\documentclass[prd,twocolumn]{revtex4}
\usepackage{graphicx, epsfig}
\usepackage{color}
\usepackage{mathrsfs}
\usepackage{bm}

\newcommand{\be}{\begin{equation}}
\newcommand{\ee}{\end{equation}}
\newcommand{\bea}{\begin{eqnarray}}
\newcommand{\eea}{\end{eqnarray}}
\newcommand{\ba}{\begin{eqnarray}}
\newcommand{\ea}{\end{eqnarray}}

\newcommand{\gapp}{\mathrel{\raise.3ex\hbox{$>$}\mkern-14mu
              \lower0.6ex\hbox{$\sim$}}}
\newcommand{\lapp}{\mathrel{\raise.3ex\hbox{$<$}\mkern-14mu
              \lower0.6ex\hbox{$\sim$}}}

\newcommand{\sothree}{${\rm SO(3)}$ }

\begin{document}
\title{Creation of Magnetic Monopoles in Classical Scattering}
\author{Tanmay Vachaspati}
\affiliation{
Physics Department, Arizona State University, Tempe, AZ 85287, USA.
}

\begin{abstract}
\noindent
We consider the creation of 't\,Hooft-Polyakov magnetic monopoles by scattering 
classical wave packets of gauge fields. An example with eight clearly separated magnetic 
poles created with parity violating helical initial conditions is shown. No clear separation of
topological charge is observed with corresponding parity symmetric initial conditions.
\end{abstract}

\maketitle

Magnetic monopoles are of key interest in current research as they embody non-perturbative 
aspects of field theories. Their rich physical and mathematical properties have inspired continued 
investigations ever since Dirac first proposed their existence
({\it e.g.} \cite{Rebbi:1985wg,Preskill:1986kp,Vilenkin:2000jqa,Manton:2004tk}). 
Dualities that relate the spectra of particles and magnetic monopoles can be an important
element in solving strongly coupled problems \cite{Goddard:1976qe,Seiberg:1994rs}
and may 
also help understand the spectrum of fundamental particles \cite{Vachaspati:1995yp,Liu:1996ea}
In particle physics, monopoles necessarily arise in grand unified models of particle 
physics, and the standard electroweak model contains field configurations that correspond to 
confined monopoles \cite{Achucarro:1999it}.

The current investigation involves the interpretation of magnetic monopoles in terms of 
particles.
Can we create magnetic monopoles by assembling particles? This problem is difficult
because particles are the quanta in a {\it quantum} field theory and magnetic monopoles
are classical objects in that field theory. No perturbative expansion of the quantum 
field theory in powers
of coupling constants can describe magnetic monopoles because properties of the
magnetic monopole are proportional to {\it inverse} powers of the coupling constant.
(Recent work on resurgence in quantum mechanics \cite{Dunne:2012zk} offers a glimmer 
of hope that divergences in the perturbative expansion may hold non-perturbative
information.)
A more modest objective is to study the creation of magnetic monopoles by scattering
{\it classical} waves, where the classical waves can themselves be thought of as 
quantum states containing high occupation numbers of quanta. 
This is the approach we shall take.

Past work on the creation of kinks in 1+1 dimensions 
\cite{Dutta:2008jt,Romanczukiewicz:2010eg,Demidov:2011dk,Demidov:2011eu,Vachaspati:2011ad,
Lamm:2013ye,Demidov:2015nea}, on the decay of 
electroweak sphalerons \cite{Copi:2008he,Chu:2011tx}, and on the scattering and annihilation of 
magnetic monopole-antimonopole \cite{Vachaspati:2015ahr}, together with results from 
magneto-hydrodynamics (MHD) \cite{1998pfp..book.....C},
offers some guidance on initial conditions that may be suitable for
creating magnetic monopoles. We will further explain these motivations when 
describing our initial conditions.

We will work with an SO(3) field theory, as considered by 't Hooft \cite{'tHooft:1974qc} and
Polyakov \cite{Polyakov:1974ek}, that contains
a scalar field in the adjoint representation, $\phi^a$ ($a=1,2,3$), and gauge fields, 
$W_\mu^a$, with the Lagrangian
\begin{equation}
L = \frac{1}{2} (D_\mu \phi)^a (D^\mu \phi)^a - \frac{1}{4} W^a_{\mu\nu} W^{a \mu\nu}
             - \frac{\lambda}{4} ( \phi^a \phi^a - \eta^2 )^2
\label{lagrangian}
\end{equation}
where,
\begin{equation}
(D_\mu \phi)^a = \partial_\mu\phi^a - i g W^c_\mu (T^c)^{ab} \phi^b 
\end{equation}
and the \sothree generators are $(T^a)^{bc} = -i \epsilon^{abc}$. The gauge field strengths
are defined by
\begin{equation}
W_ {\mu \nu}^a = \partial_\mu W_ \nu^a - \partial_ \nu W_\mu^a + 
  g \epsilon^{abc} W_\mu^b W_ \nu^c.
\label{Wmunu}
\end{equation}

Our numerical methods are borrowed from Numerical Relativity \cite{2010nure.book.....B}.
We use temporal gauge $W^a_0=0$ and treat $\Gamma^a \equiv \partial_i W^a_i$ as new 
variables whose evolution ensures that the Gauss constraints are satisfied.
The resulting classical equations of motion that we want to solve are written as
\begin{eqnarray}
\partial_t^2 \phi^a &=& \nabla^2 \phi^a
 - g \epsilon^{abc}\partial_i\phi^b W_i^c- g \epsilon^{abc} (D_i \phi)^b W_i^c \nonumber \\
&& \hskip 0.25 in
- \lambda (\phi^b\phi^b-\eta^2)\phi^a- g \epsilon^{abc} \phi^b \Gamma^c
\end{eqnarray}
\begin{eqnarray}
\partial_t W^a_{0i} &=& \nabla^2 W^a_i
+ g \epsilon^{abc} W^b_j \partial_j W^c_i - g\epsilon^{abc} W^b_j W^c_{ij} \nonumber \\
&&
- D_i \Gamma^a - g\epsilon^{abc} \phi^b (D_i\phi)^c
\end{eqnarray}
\begin{eqnarray}
\partial_t \Gamma^a &=& \partial_i W^a_{0i}
- g_p^2 [ \partial_i (W^a_{0i})  +g\epsilon^{abc}W^b_i W^c_{0i} \nonumber \\
&& \hskip 1 in
+ g\epsilon^{abc}\phi^b (D_t \phi)^c ]
\label{Gammaeq}
\end{eqnarray}
where $W^a_{0i} = \partial_t W^a_i$ in the temporal gauge,
$D_i\Gamma^a \equiv \partial_i\Gamma^a-g\epsilon^{abc}\Gamma^bW^c_i$,
and $g_p^2$ is a free parameter. Analytically, the square bracket in Eq.~(\ref{Gammaeq})
vanishes due to the Gauss constraints and the value of $g_p^2$ is irrelevant.
However the square bracket does not vanish when we discretize
the system and a non-zero value of $g_p^2$ is critical to ensure numerical stability 
\cite{2010nure.book.....B}. After some experimentation we set $g_p^2=0.75$ in our runs.
We also set $g=0.5$, $\lambda=1$ and $\eta=1$ in our numerical work.

The fields are evolved using the explicit Crank-Nicholson method with two iterations
\cite{Teukolsky:1999rm}.
We have used a new implementation of absorbing boundary conditions. Essentially, only 
the Laplacian of the fields on the lattice boundaries are replaced using radially outgoing 
boundary conditions. For example,
\begin{equation}
\nabla^2 \phi^a \to -{\hat r} \cdot \nabla (\partial_t \phi^a )
\end{equation}
at a boundary point with ${\hat r}$ the unit radial vector from the center of the box. The first
order spatial derivatives throughout the equations of motion are evaluated using one-sided
differences. We have found good stability and smooth evolution with this strategy.

The non-algorithmic part of this project is to devise initial conditions that are likely
to result in monopole creation. As noted in Ref.~\cite{Vachaspati:2011ad},
a crucial hint comes from the conservation of helicity in MHD
in plasmas with high electrical conductivity. (Magnetic helicity is 
defined as the volume integral of ${\bm A}\cdot {\bm B}$
where ${\bm A}$ is the electromagnetic gauge potential and ${\bm B}=\nabla \times {\bm A}$.)
Combined with the observed conservation of electromagnetic helicity during sphaleron decay 
\cite{Copi:2008he,Chu:2011tx} and the repulsive 
force between monopoles and antimonopoles that are twisted and that yield 
magnetic helicity on
annihilation \cite{Vachaspati:2015ahr}, it seems like a good idea to try initial conditions that 
are built from
helical, {\it i.e.} circularly polarized gauge waves. Also, MHD simulations indicate
that helicity causes magnetic fields to expand out to larger length scales (``inverse
cascade''), so that by colliding helical waves, helicity will get compressed, causing
tension against the natural tendency to expand. This tension
can relax if helicity conservation is violated, either with a decrease in the plasma 
electrical conductivity or by producing magnetic monopoles.

The natural way to discuss initial conditions is to first specify the value of the
scalar field since this determines the massless and massive components of the
gauge fields. In the numerics, however, it is easier to specify the gauge field
and then make various choices for the uniform value of the scalar field, and
this is how we will present the initial conditions.

We choose only one of the 3 SO(3) gauge fields to be non-trivial in the
initial conditions. Let this be $W^3_i$. 
Initially, at $t=0$, $W^3_i$ is given separately for
waves propagating in the $+z$ and $-z$ direction in terms of
scalar functions $f_1(x,y)$, $f_2 (t+(z-z_0))$ and $f_3  (t-(z+z_0))$ with $z_0 > 0$.
For the waves that are functions of $t+(z-z_0)$, we have:
\begin{eqnarray}
 W^3_x &=& \partial_y f_1 (\omega  f_2  - \partial_z f_2 ) \cos(\omega (t+(z-z_0))) \\
 W^3_y &=& \partial_x f_1 (\omega f_2  +\partial_z f_2 ) \sin(\omega (t+(z-z_0))) \\
 W^3_z &=& \partial_x\partial_y f_1  f_2  [\cos(\omega (t+(z-z_0))) \nonumber \\
 && \hskip 1 in 
 -\sin(\omega (t+(z-z_0)))]
\end{eqnarray}
In this form it is easy to see that $\nabla\cdot {\bm W}^3 =0$. Then 
$\partial_t W^3_i = +\partial_z W^3_i$ gives
\begin{eqnarray}
\partial_t  W^3_x &=& \partial_y f_1 [ (\omega \partial_z f_2  - \partial_z^2 f_2 )
 \cos(\omega (t+(z-z_0)) \nonumber \\
 && \hskip 0.3 in 
 - (\omega  f_2 -\partial_z f_2 )\omega \sin(\omega (t+(z-z_0))]\\ 
\partial_t  W^3_y &=&   \partial_x f_1 [(\omega\partial_z f_2 +\partial_z^2 f_2 )
 \sin(\omega (t+(z-z_0)) \nonumber \\
 && \hskip 0.3 in
 + (\omega f_2  +\partial_z f_2 )\omega \cos(\omega (t+(z-z_0))]\\
\partial_t  W^3_z &=& \nonumber \\
&& \hskip -1.4 cm
\partial_x\partial_y f_1 [  \partial_z  f_2   (\cos(\omega (t+(z-z_0))-\sin(\omega (t+(z-z_0))) \nonumber \\
&& \hskip -1.4 cm
+ \omega  f_2  (-\sin(\omega (t+(z-z_0)) -\cos(\omega (t+(z-z_0)))]
\end{eqnarray}
Since $\nabla \cdot {\bm W}^3=0$, and the electric field ${\bm E}^3 = -\partial_t {\bm W}^3$,
the Gauss constraint is satisfied with vanishing charge density. We will arrange
for a vanishing charge density by taking the scalar field to have vanishing
time derivative initially 
\begin{equation}
\partial_t \phi^a |_{t=0} =0 .
\end{equation}
We will also take $\phi^a ={\rm constant}$ initially, with different choices for the constant
describing different physical situations as discussed below.

For a packet traveling in the opposite direction, we write the formulae
in terms of $ f_3  (t-(z+z_0))$:
\begin{eqnarray}
 W^3_x &=& \partial_y f_1 (-\omega '  f_3 - \partial_z f_3 )\cos (\omega' (t-(z+z_0)) \\
 W^3_y &=& - \partial_x f_1 (\omega ' f_3 -\partial_z f_3 )\sin(\omega '(t-(z+z_0)) \\
 W^3_z &=& \partial_x\partial_y f_1  f_3  ~ (\cos(\omega' (t-(z+z_0))) \nonumber \\
 && \hskip 1 in
 -\sin(\omega ' (t-(z+z_0)))
\end{eqnarray}
For these packets we use $\partial_t W^3_i = -\partial_z W^3_i$ to write
\begin{eqnarray}
\partial_t  W^3_x &=& -\partial_y f_1 [
(-\omega ' \partial_z f_3  - \partial_z^2 f_3 )\cos(\omega ' (t-(z+z_0)))\nonumber \\
&& \hskip 0.1 in
- (\omega ' f_3 +\partial_z f_3 )\omega ' \sin(\omega ' (t- (z+z_0))] \\
\partial_t  W^3_y &=&
 \partial_x f_1 [
(\omega ' \partial_z f_3 -\partial_z^2 f_3 )\sin(\omega ' (t-(z+z_0))) \nonumber \\
&& \hskip 0.1 in
- (\omega ' f_3 -\partial_z f_3 )\omega ' \cos(\omega ' (t-(z+z_0))) ]\\
\partial_t  W^3_z &=& 
- \partial_x\partial_y f_1 [\partial_z f_3  (\cos(\omega ' (t-(z+z_0))) \nonumber \\
&& \hskip 1 in
-\sin(\omega ' (t-(z+z_0)))) \nonumber \\
&& \hskip -1.5 cm
+\omega ' f_3  (\cos(\omega ' (t-(z+z_0)))+\sin(\omega ' (t-(z+z_0)))) ]
\end{eqnarray}

The profile functions are taken such as to create a localized packet in all directions
\begin{equation}
f_1(x,y) = a~\exp \left [ - \frac{(x^2+y^2)}{2w^2} \right ]
\end{equation}
\begin{equation}
f_2(t+(z-z_0)) = \exp \left [ - \frac{(t+(z-z_0))^2}{2w^2} \right ]
\end{equation}
\begin{equation}
f_3(t-(z+z_0)) = \exp \left [ - \frac{(t-(z+z_0))^2}{2w^2} \right ]
\end{equation}
where $a$ is an amplitude and $w$ is a width.
The frequencies $\omega$ and $\omega '$ can be different in general but 
we only consider $\omega' = \pm \omega$. The case $\omega '=\omega$
corresponds to scattering of left- and right-handed circular polarizations, while
$\omega'=-\omega < 0$ corresponds to scattering of left- on left-handed 
circular polarization waves. 

Now we linearly superpose the counterpropagating wave packets and set 
$t=0$ to get the initial conditions for the gauge fields for our scattering experiments.

Next we discuss the choice of the scalar field $\phi^a$. The simplest choice is
$\phi^1=0=\phi^2$, $\phi^3=\eta$ but this is too simple. In this case, ${\bm W}^3$ corresponds
to the massless ``photon'' of the model, and in this classical evolution, the scattering of photons
does not excite any other field. In other words, the dynamics lies in a subspace of
the full field theory \cite{Barriola:1993fy} and the classical dynamics is exactly as it would be
in Maxwell theory. The next choice we considered is $\phi^1=\eta$, $\phi^2=0=\phi^3$.
Now ${\bm W}^3$  is a massive boson of the theory. This too leads to dynamics in a subspace,
namely that spanned by \{$\phi^1$, $\phi^2$,${\bm W}^3$\}. So now the model is effectively
the Abelian-Higgs U(1) model. It is interesting that when we performed some runs with these
initial conditions, we did observe zeros of $\phi^a$, suggesting that we had created loops of
strings. We will postpone this investigation for the future since here we are focusing
on the production of magnetic monopoles.

For the classical dynamics to explore the full model, we take 
\begin{equation}
\phi^1 =\frac{\eta}{\sqrt{2}}, \ \phi^2=0, \ \phi^3=\frac{\eta}{\sqrt{2}}
\end{equation}
at $t=0$. Now the initial gauge field wave packet is a superposition of the photon and the 
massive gauge boson.

After the system has evolved for a while, we would like to know if monopoles have
been created. Since monopoles are stable objects and the scalar field vanishes
at their centers, the existence of a monopole can be detected by
looking for peaks of the potential energy density that are close to the value
$\lambda \eta^4/4 =0.25$. We follow the potential energy diagnostic with a
calculation of the topological winding which is defined as
\begin{equation}
W(S) = 
\frac{1}{8\pi} \oint_S d{\hat n}^i 
\epsilon_{ijk}\epsilon_{abc} {\hat \phi}^a \partial_j{\hat \phi}^b\partial_k{\hat \phi}^c 
\end{equation}
where ${\hat n}$ is the outward unit normal to a closed surface $S$ and
${\hat \phi}^a=\phi^a/|{\vec \phi}|$.
We replace the continuum formula for the winding with a discrete formula as follows.
We first define the vector, ${\vec v}$, at every vertex of the lattice,
\begin{equation}
v_i = \epsilon_{ijk}\epsilon_{abc} {\hat \phi}^a \partial_j{\hat \phi}^b\partial_k{\hat \phi}^c.
\label{vivec}
\end{equation}
Then the winding for a fundamental cell of the simulation lattice is given by
\begin{eqnarray}
{\overline W}(S) = \frac{1}{8\pi} \sum_{\rm plaq.} \left ( \frac{1}{4} \sum_{\rm vertices} {\hat n}^i v_i \right )
\end{eqnarray}
where the outside sum is over the 6 plaquettes bounding a cell, ${\hat n}$ is the unit
vector normal to the plaquette, $v_i$ is the vector in Eq.~(\ref{vivec}) evaluated
at the vertices of the plaquette, and the $1/4$ is due to an averaging over the
4 corners of the plaquette. 
Even though $W(S)$ takes integer values, the discrete version ${\overline W}(S)$ may
not be an integer. However, for large surfaces $S$, ${\overline W}(S)$ will also tend
towards an integer value.

Our simulations are run on a $128^3$ lattice with lattice spacing $dx=0.1$ with 
field theory parameters: $g=0.5$, $\lambda=1$, $\eta=1$. The 
initial condition parameters were chosen to be: $w=0.4$, $z_0=1$, $a=10$, 
$\omega=4$. With this choice of parameters, the 
initial energy is $\sim 10^5$ and is much larger than the energy per monopole-antimonopole
pair, which is $\sim 10^2$. Further exploration of parameters and choice of initial 
conditions is likely to yield monopoles even when we start with less energy, though
intuitively the initial conditions will have to be more finely tuned or ``coherent'' 
if we take lower initial energy. 

\begin{figure}
  \includegraphics[height=0.25\textwidth,angle=0]{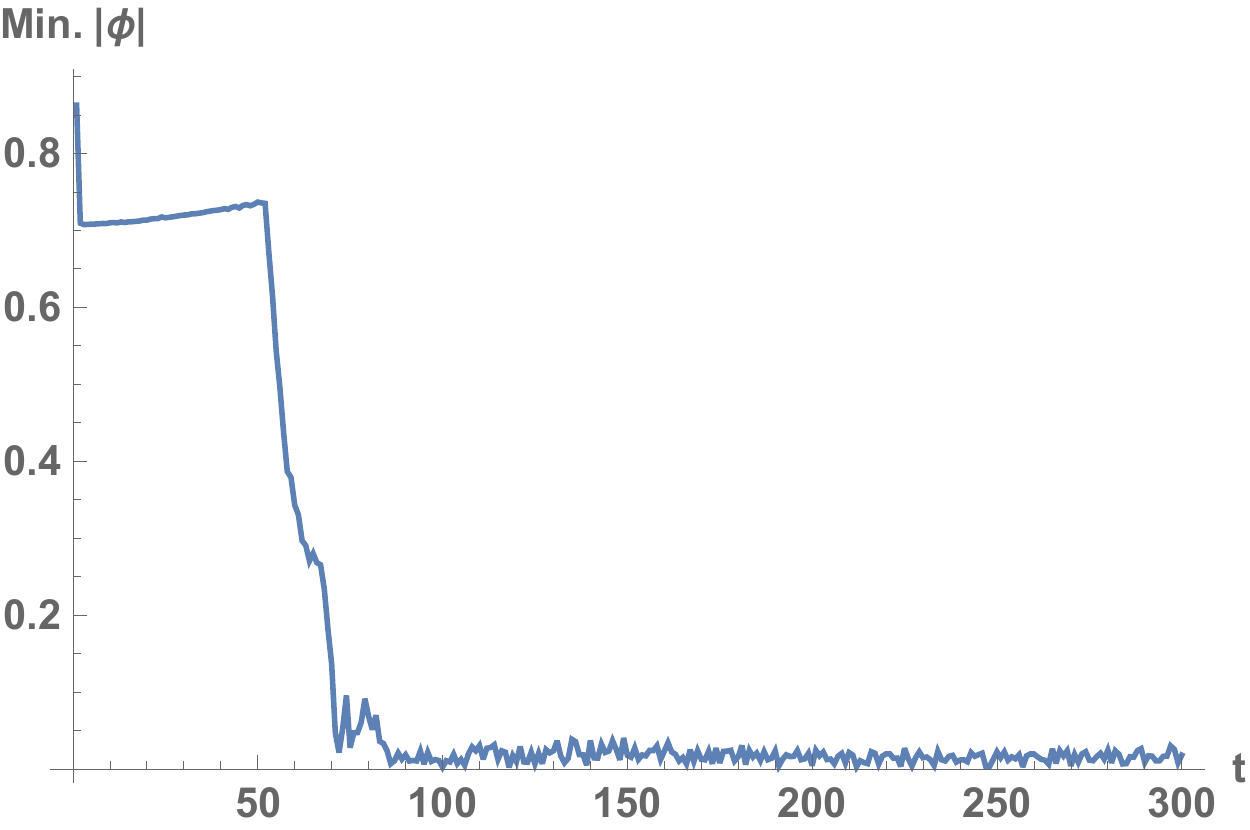}
  \caption{Minimum value of $|{\vec \phi}|$ on the lattice as a function of time showing
  that zeros of the scalar field are produced after some evolution.}
\label{phimin}
\end{figure}

The first indication that monopoles have been produced during evolution
is that we see zeros of the Higgs. This is shown in Fig.~\ref{phimin}.

\begin{figure}
  \includegraphics[height=0.25\textwidth,angle=0]{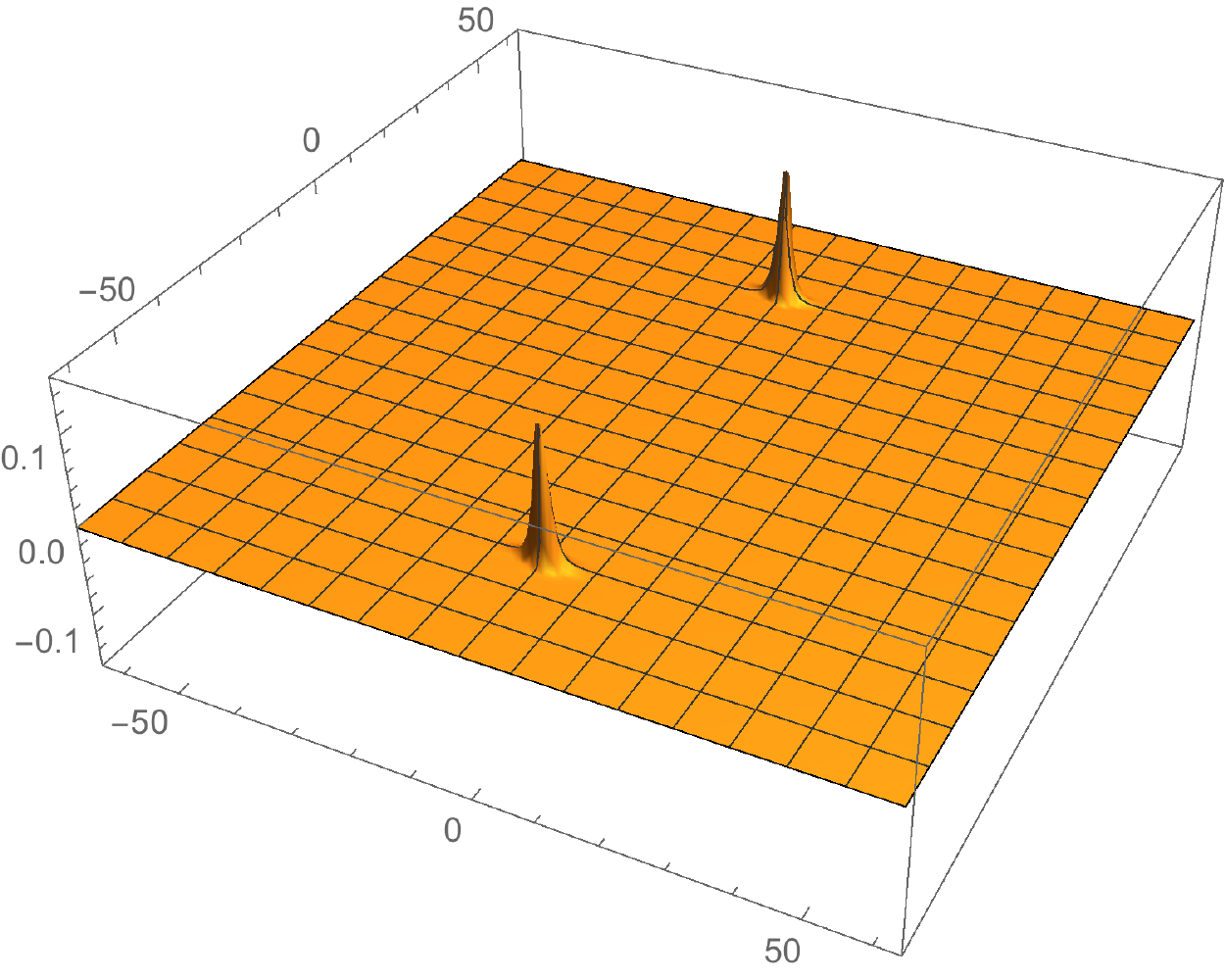}
   \includegraphics[height=0.25\textwidth,angle=0]{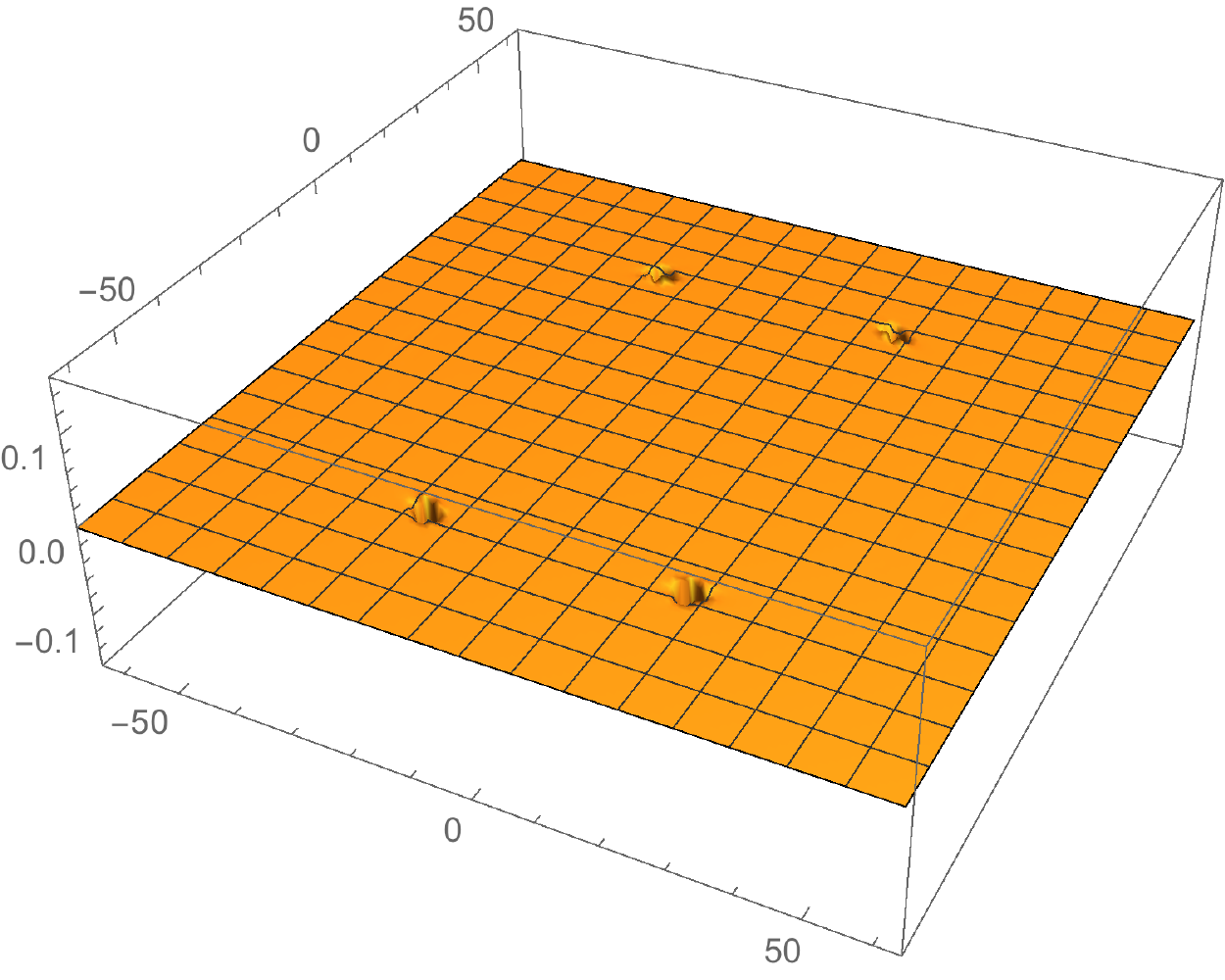}
    \includegraphics[height=0.25\textwidth,angle=0]{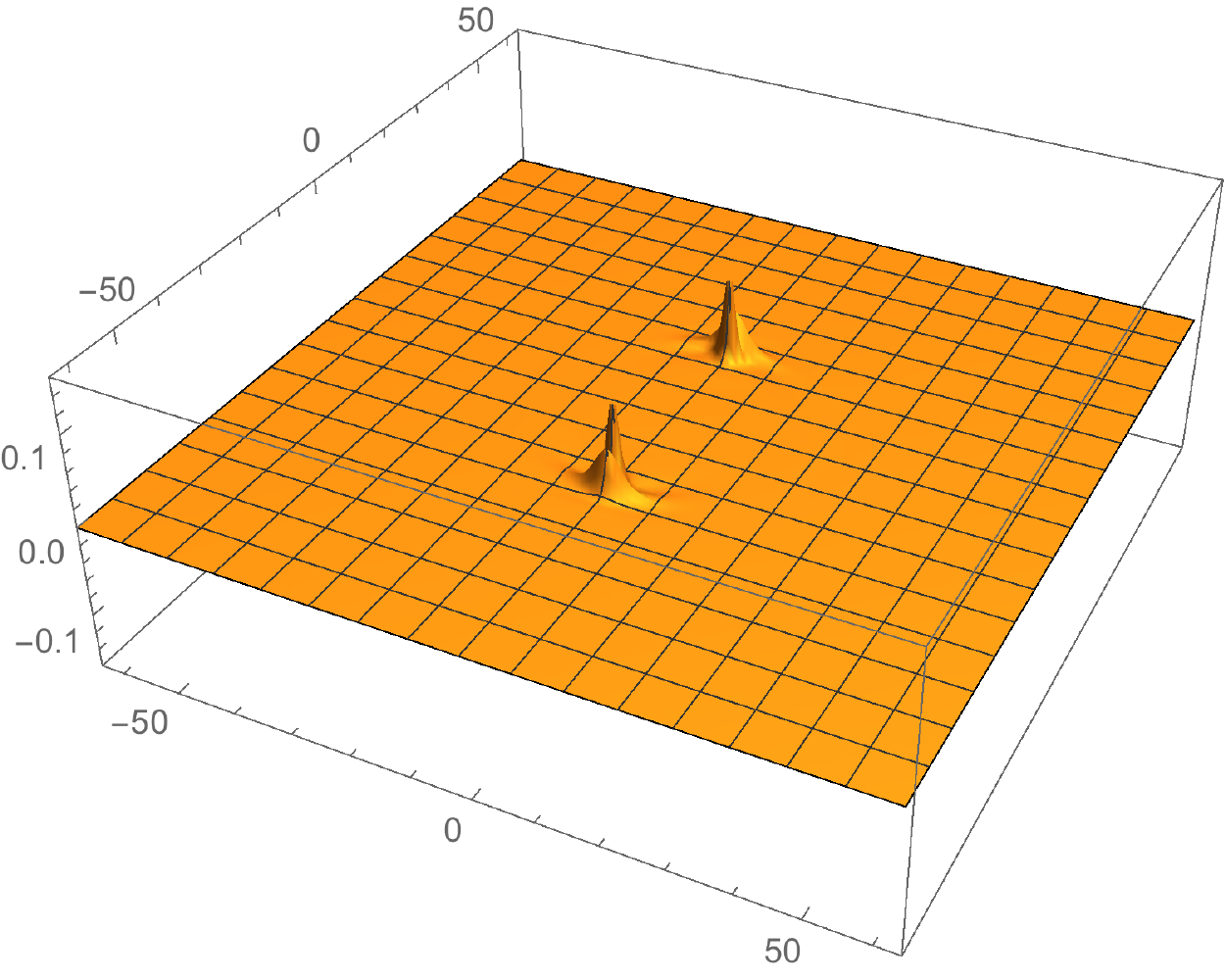}
  \caption{Topological winding at late times on slices with $z=9.2, 10.1$ and $12.1$.
  The total topological charges on these slices are +2, -4, and +2 respectively.}
\label{windingonslices}
\end{figure}

\begin{figure}
  \includegraphics[height=0.2\textwidth,angle=0]{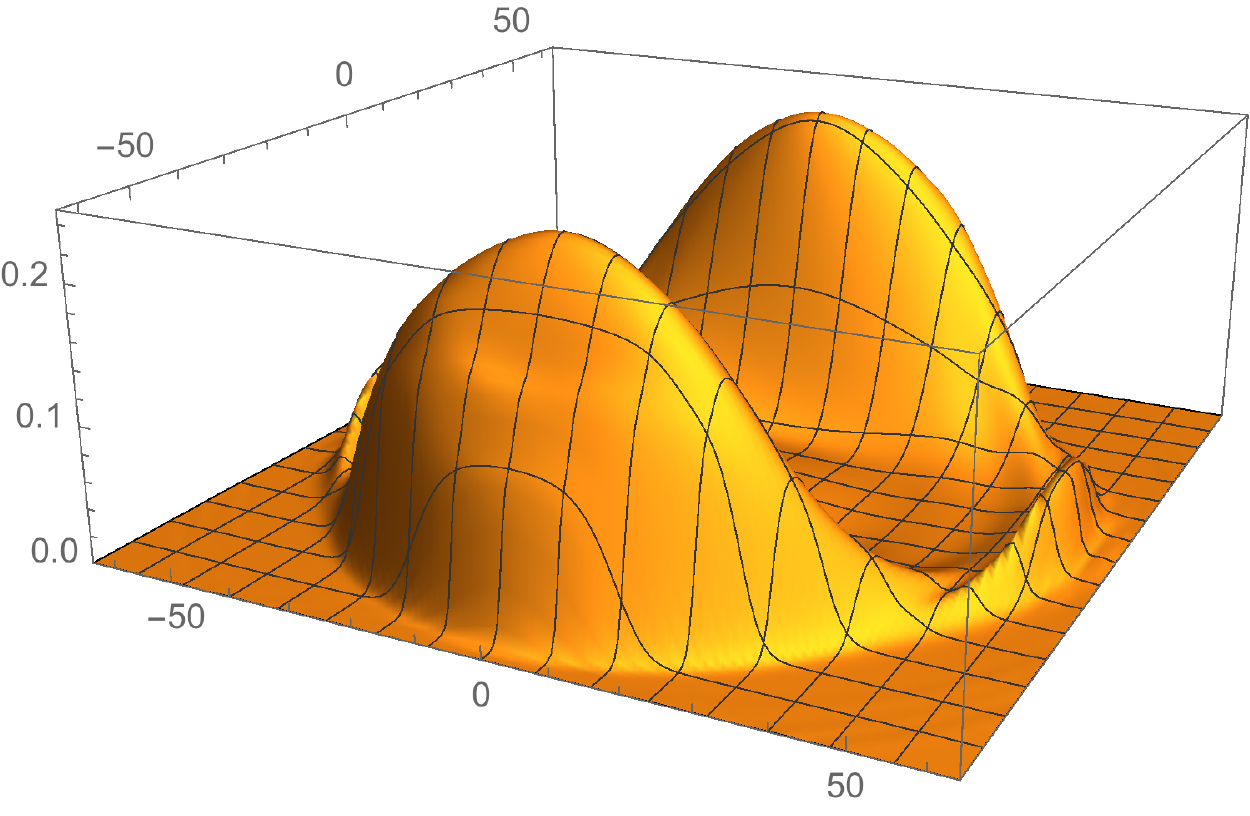}
   \includegraphics[height=0.2\textwidth,angle=0]{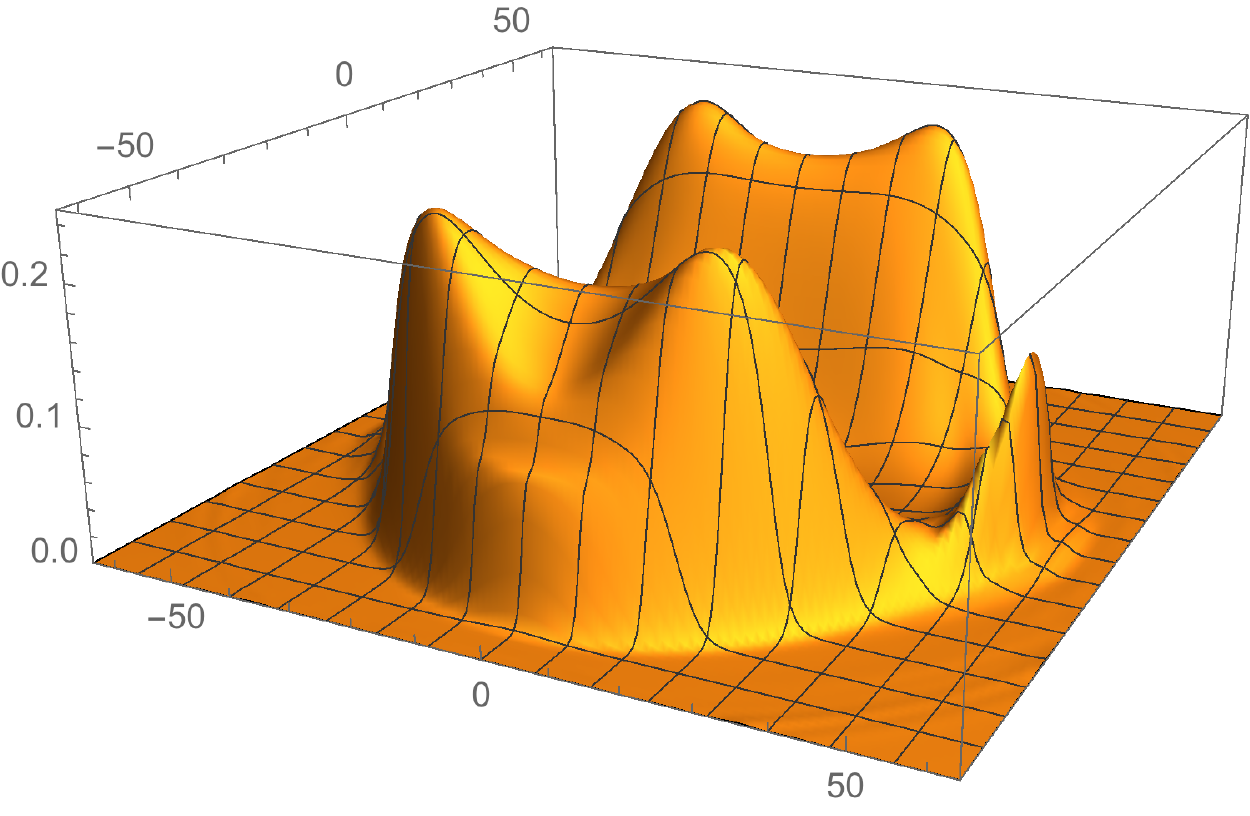}
    \includegraphics[height=0.2\textwidth,angle=0]{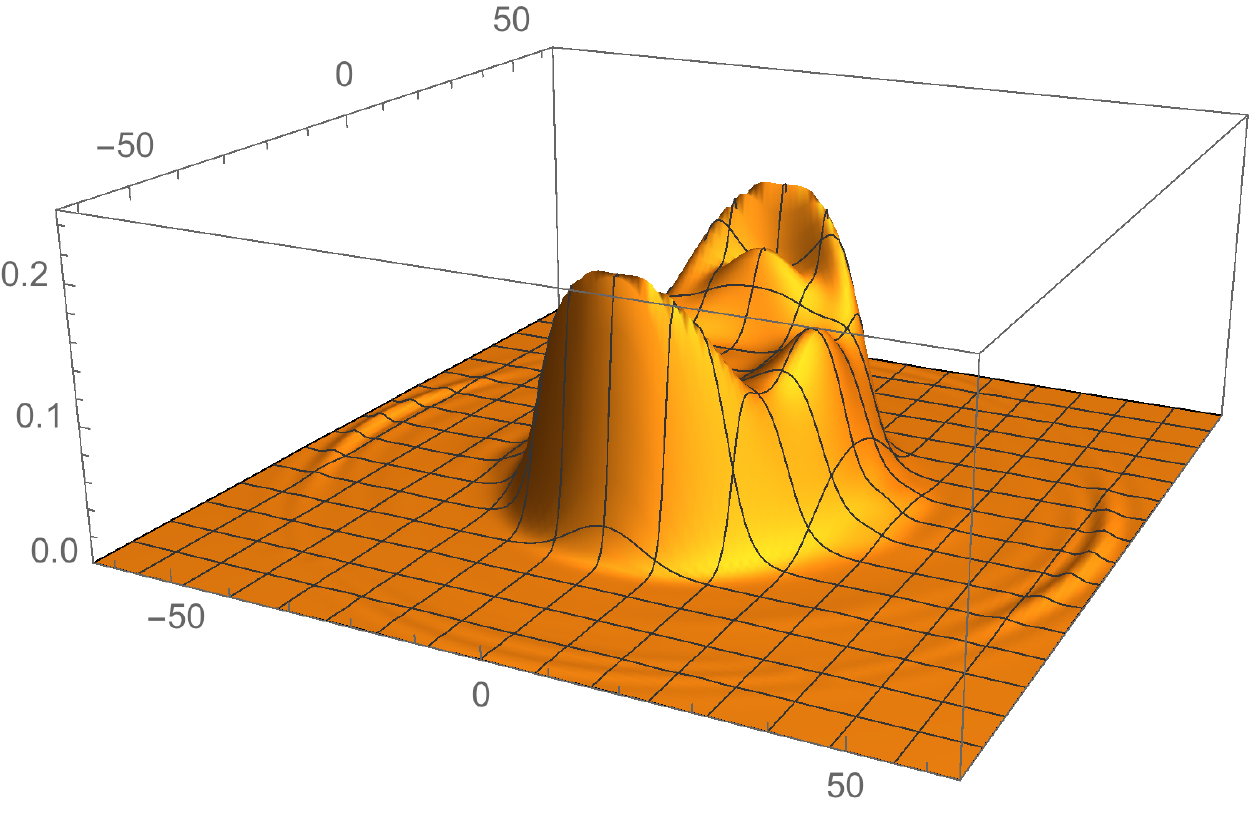}
  \caption{Potential energy density distribution at the final time of the simulation
  on spatial slices with $z=9.2, 10.1$ and $12.1$. With $|{\vec \phi}|=0$, the
  potential energy density is 0.25 for our parameters.}
\label{potenergyonslices}
\end{figure}

The presence of monopoles is confirmed by finding the topological
winding, ${\overline W}$ for every cell of the lattice. In Fig.~\ref{windingonslices}
we show the distribution of topological charge on xy-slices, {\it i.e.} on $z={\rm constant}$ 
slices of the lattice. Only slices with significant windings are shown and the total 
topological charge on the entire lattice vanishes. It is clear that the scattering
has resulted in 4 monopoles and 4 antimonopoles. This is further confirmed
by plotting the potential energy density on these slices, shown in 
Fig.~\ref{potenergyonslices}. The peaks in the potential energy represent
monopoles within which the scalar field has a zero. In the discrete simulation,
the zero may lie within a cell of the lattice and the potential will not quite be
its maximal value of 0.25.

The distances between monopoles and antimonopoles can be estimated
and is on the order of 3 monopole widths where we take the monopole
width to be the inverse scalar boson mass, $m_S = \sqrt{2\lambda}  \eta$.
We can estimate the velocities of the monopoles from Fig.~\ref{phimin}
and our choice of time step $dt=dx/4$ where $dx$ is the lattice spacing.
We find that the monopoles are relativistic with $v \sim 1$. A simple estimate 
of the monopole-antimonopole escape velocity gives $v_{\rm esc} \sim 0.1$
when the separation of the pair is a few monopole widths.
Since the monopole and
antimonopole velocities are not aligned, the monopoles and antimonopoles
are not bound and will continue to fly apart with time, as we observe
directly during the later stages of the simulation.

A curious feature of the final configuration of monopoles is that they are
all located at $z > 0$.
However, this is not in contradiction with any symmetry, since our
initial conditions for $\omega'=-\omega$ are not reflection symmetric
under $z \to -z$.

\begin{figure}
  \includegraphics[height=0.2\textwidth,angle=0]{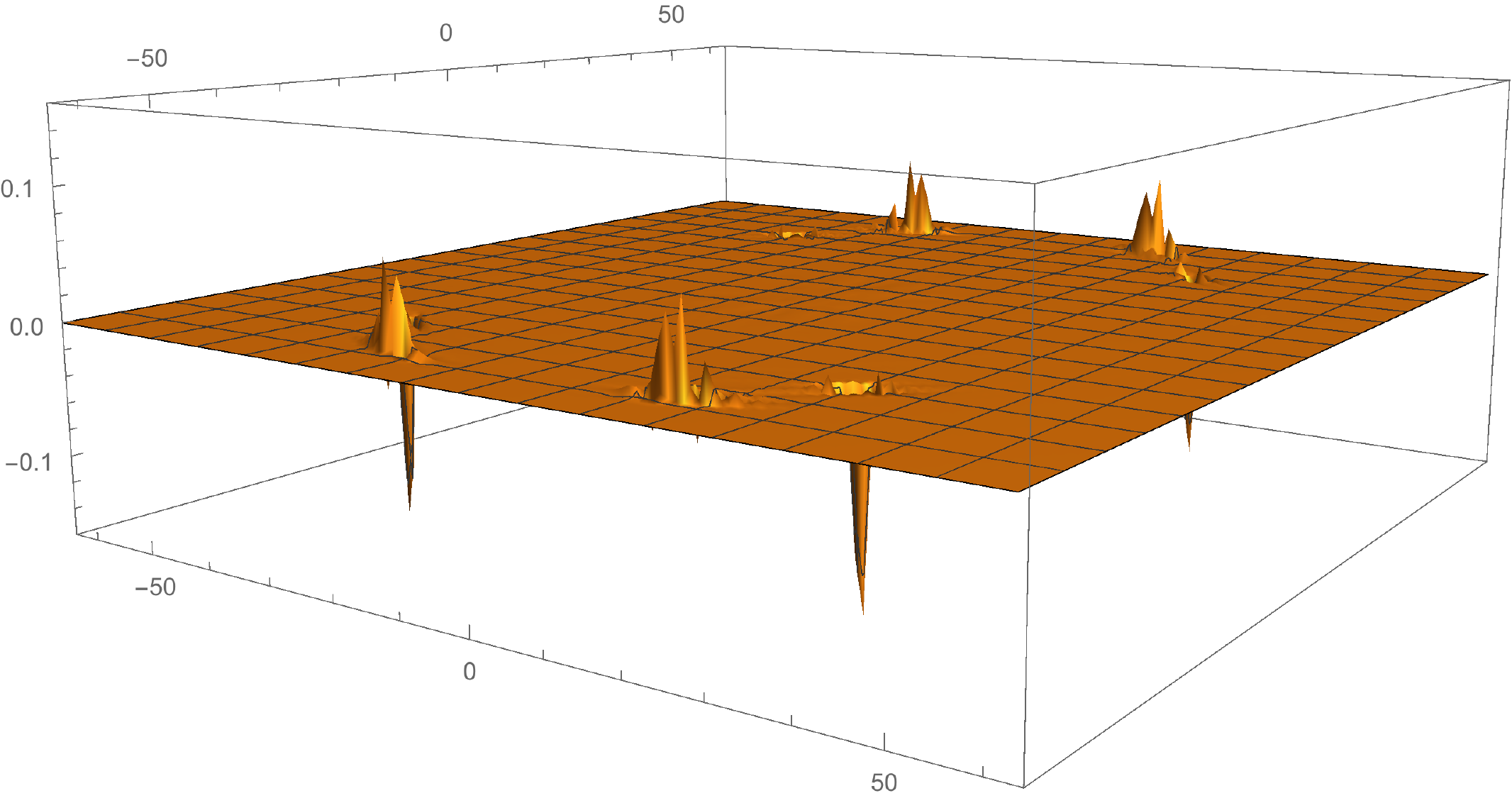}
    \includegraphics[height=0.2\textwidth,angle=0]{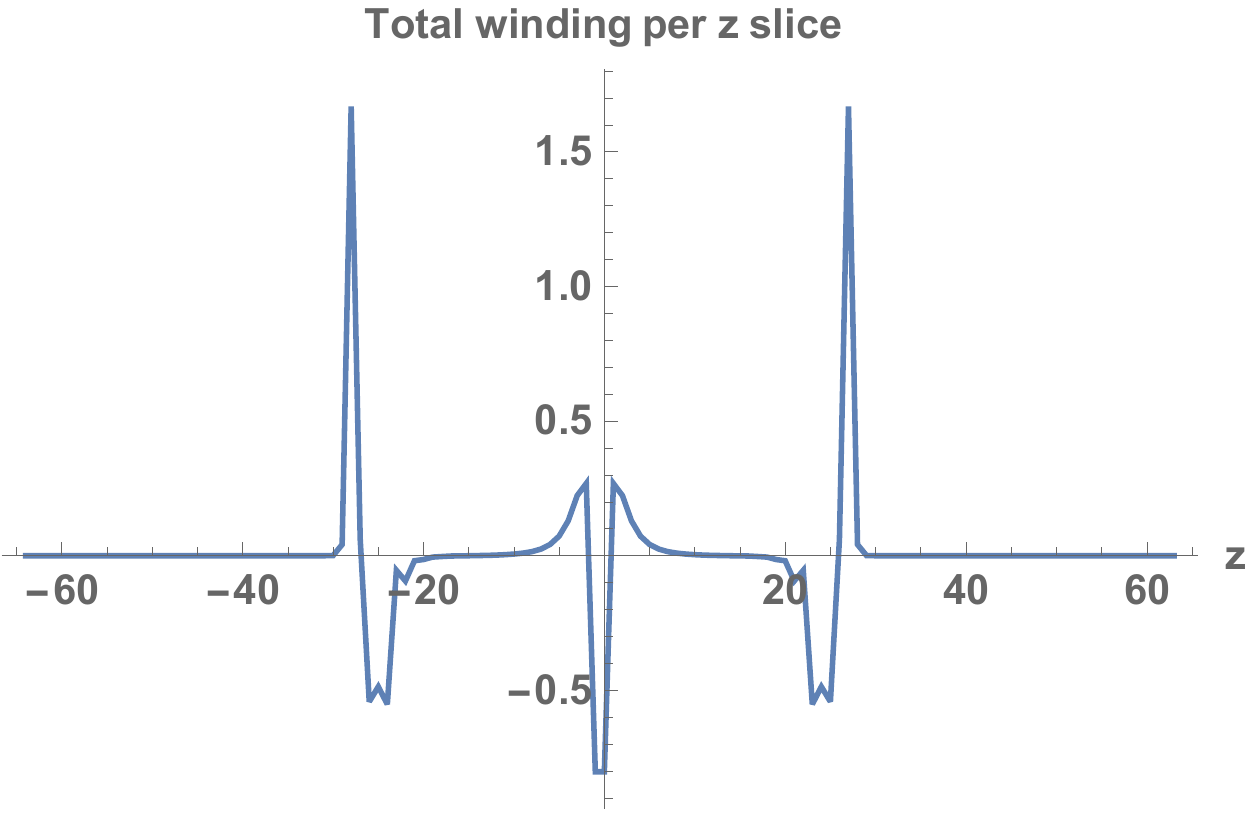}
  \caption{Topological winding on the $z=0$ slice for the $\omega=+\omega'=4$ 
  simulation. The plot does not show a clear separation of positive and negative
  winding. In the second panel we show the integrated winding per $z$ slice
  as a function of $z$. Here too we do not see a clear separation of positive
  and negative charges.}
\label{omPlus4}
\end{figure}

We intend to automate the numerical program so that it can scan over a range
of parameters and detect and record magnetic monopoles when they do occur. 
For the time being we have tried a few different values of the parameters and find
monopole creation for larger values of the amplitude $a$ and frequency $\omega$.
Of particular interest is the dependence on the sign of $\omega'$ that determines
whether we are scattering left-on right-handed waves or left- on left-handed waves.
The results discussed above are for $\omega=4$, $\omega'=-4$ (left- on left-handed
waves); so we also
ran the code with $\omega'=+4$ and all other parameters kept the same.
In Fig.~\ref{omPlus4} we show the topological winding distribution on the
$z=0$ slice. The sharp negative peaks signifying possible antimonopoles have
positive peaks in their neighborhoods and the integrated charge vanishes. There
are other peaks at non-zero $z$ but these too have canceling charge distributions
in their vicinity. The total topological charge per $z$ slice is plotted in the second panel
of Fig.~\ref{omPlus4} to further illustrate this feature. Hence, simply flipping
the handedness of one of the initial waves results in evolution in which there
is no clear separation of monopole and antimonopole charge.

The probability for creating monopoles depends on the probability measure 
on initial states and this depends on the human will to create such states. For 
example, the probability of creating a complex structure like the Large Hadron 
Collider by pure chance is incredibly low, nonetheless it exists. A more
meaningful question is the sensitivity of the outcome of the scattering to
small errors in the initial conditions. Is the creation of monopoles a ``stable''
process? In the case of kinks in 1+1 dimensions, it is known that their
scattering and annihilation is chaotic \cite{Campbell:1983xu,Anninos:1991un}. 
This ties in with the chaotic behavior seen in the creation of 
kinks \cite{Dutta:2008jt,Romanczukiewicz:2010eg}
and it appears that the creation of kinks is very sensitive to the initial conditions, 
{\it i.e.} it is unstable. However, chaos seems to be absent in the annihilation of 
magnetic monopole and antimonopole, at least within the domain of scattering
parameters that have been investigated \cite{Vachaspati:2015ahr}. This suggests 
that the creation of monopoles will also be a stable process but is something that 
needs to be investigated.

\acknowledgements

I thank Sourish Dutta, Jeff Hyde, Henry Lamm and Erick Weinberg for comments. 
I am grateful to the Institute for Advanced Study, Princeton, for hospitality during 
the course of this work. This work was also performed in part at the Aspen Center 
for Physics, which is supported by National Science Foundation grant PHY-1066293.
TV is supported by the U.S. Department of Energy, Office of High Energy Physics, 
under Award No. DE-SC0013605 at ASU.

\bibstyle{aps}
\bibliography{mmbarcreation}

\end{document}